# Estimation of Grüneisen Parameter of Layered Superconductor LaO$_{0.5}$F$_{0.5}$BiS$_{2-x}$Se$_x$ (x = 0.2, 0.4, 0.6, 0.8, 1.0)

Fysol Ibna Abbas[1,2], Kazuhisa Hoshi[1], Yuki Nakahira[1,3], Miku Yoshida[1], Aichi Yamashita[1], Hiroaki Ito[4], Akira Miura[4], Chikako Moriyoshi[5], Chul-Ho Lee[6], and Yoshikazu Mizuguchi[1*]

[1] *Department of Physics, Tokyo Metropolitan University, 1-1, Minami-osawa, Hachioji 192-0397, Japan*
[2] *Department of Electrical & Electronic Engineering, City University, Khagan, Birulia, Savar, Dhaka-1216, Bangladesh*
[3] *Quantum Beam Science Research Directorate, National Institutes for Quantum Science and Technology, Hyogo 679-5148, Japan*
[4] *Faculty of Engineering, Hokkaido University, Kita 13 Nishi 8, Sapporo 060-8628, Japan*
[5] *Graduate School of Advanced Science and Engineering, Hiroshima University, Higashihiroshima, Hiroshima, 739-8526, Japan*
[6] *National Institute of Advanced Industrial Science and Technology (AIST), Tsukuba, Ibaraki 305-8568, Japan*

The superconducting properties and structural parameters, including the Grüneisen parameter ($\gamma_G$), of the BiCh$_2$-based (Ch: S, Se) layered superconductor LaO$_{0.5}$F$_{0.5}$BiS$_{2-x}$Se$_x$ were investigated. The superconducting transition temperature ($T_c$) increased with increasing Se concentration ($x$), and bulk superconductivity was induced by Se substitution. $\gamma_G$ exhibited a larger value when $x \geq 0.4$, where the superconducting properties were improved. These results suggest that the increase of lattice anharmonicity is preferable for superconductivity in LaO$_{0.5}$F$_{0.5}$BiS$_{2-x}$Se$_x$. The discussion includes data for related superconductors REO$_{0.5}$F$_{0.5}$BiS$_2$ (RE = Pr, Nd), where a positive correlation between $\gamma_G$ and $T_c$ is revealed for Se-poor samples. In addition, specific heat analyses suggest the presence of a low-energy optical phonon mode for LaO$_{0.5}$F$_{0.5}$BiS$_{2-x}$Se$_x$.

## 1. Introduction

Following the discovery of superconductivity in Bi$_4$O$_4$S$_3$ and LaO$_{0.5}$F$_{0.5}$BiS$_2$ in 2012, BiCh$_2$-based (Ch = S, Se) layered superconductors have since been developed [1–10]. Despite the highest recorded transition temperature ($T_c$) in the BiCh$_2$-based family being as high as 11 K, exotic superconducting properties still attract researchers in the field of superconductivity science. An example is the suppression of local in-plane structural disorder in the induction of bulk superconductivity in REO$_{0.5}$F$_{0.5}$BiCh$_2$ (RE: rare earth) [11–15]. The suppression of local disorder is achieved by in-plane chemical pressure (CP) effects, which can be applied by isovalent substitution at RE and/or Ch sites. In addition, in $Bi_4O_4S_3$ [16,17], La(O,F)BiSSe [18–21], and Nd(O,F)BiS$_2$ [22,23], unconventional features of superconducting states have been experimentally revealed [24]. Further, theoretical studies suggest that unconventional pairings in BiCh$_2$-based systems are feasible [25–27]. Therefore, the mechanisms of superconductivity in BiCh$_2$-based systems should be elucidated.

Recently, we studied the relationship between the lattice anharmonicity in RE(O,F)BiS$_2$ (RE = La, Ce, Pr, Nd) by estimating the Grüneisen parameter ($\gamma_G$) [28], which was motivated by the study of $\gamma_G$ in a related system LaOBiS$_{2-x}$Se$_x$ [29], which is a thermoelectric system that exhibits an anharmonic atomic vibration [30–32]. In RE(O,F)BiS$_2$, a positive correlation between superconductivity ($T_c$) and $\gamma_G$ was observed [28], implying that the increase in anharmonicity correlates with the emergence of superconductivity in BiCh$_2$-based systems. This trend is similar to that observed in caged compounds with rattling modes [33,34]. Therefore, additional investigation of the relationship between superconductivity and anharmonicity in BiCh$_2$-based compounds is required to understand the superconductivity mechanisms in the superconductor family. In this work, we investigated the superconducting properties and lattice anharmonicity of LaO$_{0.5}$F$_{0.5}$BiS$_{2-x}$Se$_x$, in which Se substitution at the in-plane Ch1 site induces in-plane CP and bulk superconductivity [11,14]. The results on $\gamma_G$ suggest that bulk superconductivity is induced when anharmonicity is enhanced. However, the estimated $\gamma_G$ was relatively constant at $x$ = 0.4–1.0, which could be the cause of $T_c$ being lower in LaO$_{0.5}$F$_{0.5}$BiS$_{2-x}$Se$_x$ than in NdO$_{0.5}$F$_{0.5}$BiS$_2$ with a larger $\gamma_G$. In addition, we found that $\gamma_G$ for Se-rich samples deviates from the positive correlation in $\gamma_G$-$T_c$, which implies that the effects of in-plane chemical pressure through Se substitution are different from those of Nd substitution at the blocking layers. The presence of optical phonons with an energy of 6–7 meV that correspond to the Bi rattling modes [32] were detected from specific heat ($C$) analyses in LaO$_{0.5}$F$_{0.5}$BiS$_{2-x}$Se$_x$.

## 2. Experimental Details

*2.1 Synthesis and Measurements*

Polycrystalline samples of LaO$_{0.5}$F$_{0.5}$BiS$_{2-x}$Se$_x$ ($x$ = 0.2, 0.4, 0.6, 0.8, and 1.0) were synthesized via a solid-state reaction [35]. The obtained samples were densified using a high-pressure synthesis method with a 180-ton cubic-anvil system at a pressure of ~1.5 GPa and an annealing temperature of ~300 °C for 15 min, which allowed us to examine the sound velocity using dense samples with a relative density higher than 95%. For $x$ = 0, we analyzed low-temperature specific heat using a polycrystalline sample, and the $\gamma_G$ was estimated using the parameters published in Ref. [28]. To investigate the temperature dependence of the lattice constants, synchrotron X-ray diffraction (SXRD) experiments were performed at beamline BL02B2 and SPring-8 at a wavelength of 0.496571(1) Å under proposal No. 2022A1698. The SXRD experiments were performed with a sample rotator system at room temperature, and the diffraction data were collected using a high-resolution one-dimensional semiconductor detector, multiple MYTHEN system [36], with a step size of $2\theta = 0.006°$. The samples were packed in an evacuated capillary glass tube with an inner diameter of 0.2 mm, and the temperature of the sample was controlled using $N_2$ gas. From the obtained SXRD patterns, lattice constants $a$ and $c$ were determined by the Rietveld method using RIETAN-FP [37]. We determined the Se occupancy at the in-plane (Ch1) and out-of-plane (Ch2) sites and confirmed that Se selectively occupied the in-plane Ch1 site, as shown in Fig. 1 [38]. The refined Se concentrations are listed in Table I. A schematic of the crystal structure (Fig. 1) was obtained using VESTA [39]. Magnetic susceptibility ($\chi$) was measured using a superconducting interference device (SQUID) magnetometer on a Magnetic Property Measurement System (MPMS3, Quantum Design) with an applied field of 10 Oe after zero-field cooling and field cooling (ZFC and FC). Specific heat was measured via a relaxation method using a physical property measurement system (PPMS, Quantum Design). The longitudinal sound velocity ($v_L$) of the sample was measured using an ultrasonic thickness detector (Satotech-2000C). The actual composition was investigated by energy-dispersive X-ray spectroscopy (EDX) using a scanning electron microscope (TM-3030, Hitachi-High Tech) equipped with an EDX analyzer SwiftED (Oxford).

2.2 Grüneisen Parameter

The $\gamma_G$ values of the samples were calculated using the following formula, as described in our previous studies [28,29]:

$$\gamma_G = \frac{\beta_V B V_{\text{mol}}}{C_V} \qquad (1)$$

where $\beta_V$, $B$, $V_{\text{mol}}$, and $C_V$ are the volume thermal-expansion coefficient, bulk modulus, molar volume, and specific heat, respectively. The parameters needed for the estimation of $\gamma_G$ were calculated using equations (2–5, where $dV/dT$, $\rho$, $v_L$, $v_S$, $v_m$, $\theta_D$, $h$, $k_B$, $n$, $N_A$, and $M$ denote the temperature gradient of the lattice volume, density of the material, longitudinal sound velocity, shear sound velocity, average sound velocity, Debye temperature, Plank's constant, Boltzmann's constant, number of atoms in the molecule (formula unit), Avogadro's constant, and molecular weight (per formula unit), respectively.

$$\beta_V = \frac{1}{V(300\text{ K})}\frac{dV}{dT} \qquad (2)$$

$$B = \rho\left(v_L^2 - \frac{4}{3}v_S^2\right) \qquad (3)$$

$$\theta_D = \left(\frac{h}{k_B}\right)\left[\frac{3n}{4\pi}\left(\frac{N_A\rho}{M}\right)\right]^{\frac{1}{3}}v_m \qquad (4)$$

$$v_m = \left[\frac{1}{3}\left(\frac{2}{v_S^3} + \frac{1}{v_L^3}\right)\right]^{-\frac{1}{3}} \qquad (5)$$

## 3. Results

### 3-1. Superconducting Properties

The superconducting properties of the obtained samples were investigated using magnetic susceptibility ($\chi$) and specific heat ($C$). Figure 2(a) shows the temperature dependence of $4\pi\chi$ for $x = 0.2$–1.0. Figure 2(b) shows the temperature dependence of $4\pi\chi$ for $x = 0$, where filamentary signals of superconductivity was observed. At $x = 0.2$, the shielding volume fraction examined by ZFC data at 1.8 K was less than -1, but it increased with increasing $x$. Here, the shielding volume fraction exceeded -1 owing to the demagnetization effect for samples with bulk superconductivity, which has been commonly observed in BiCh$_2$-based polycrystalline samples [10]. The trend of Se substitution inducing bulk superconductivity is consistent with the results of a previous study [35]. The $T_c$ values estimated from the temperature at which $\chi$ begins to decrease are plotted in Fig. 5(a). Figs. 2(c–h) show the temperature dependences of $C_{el}/T$, where $C_{el}$ is the electronic specific heat estimated by subtracting the phonon contribution of $C$. We assumed that the low-temperature $C$ is described as $C = \gamma T + \beta T^3 + \delta T^5$ [40,41]. $T_c$ in the specific heat measurements was estimated from the analysis of the jump of $C_{el}$ at $T_c$ ($\Delta C_{el}$), where the entropy balance was considered in the analyses, as shown in Figs. 2(c–h). The trend of the Se concentration dependence of $T_c$ in the specific heat data is similar to that in the susceptibility data. The $\Delta C_{el}/\gamma T_c$ value for $x = 0$ is the smallest, which is consistent with the susceptibility result. The $\Delta C_{el}/\gamma T_c$ values were comparable at $x = 0.2, 0.4$, and $0.6$, and increased at $x = 0.8$ and $1.0$. The results suggest that the bulk nature of superconductivity is enhanced by Se substitution because the estimated value at $x = 1.0$ is close to that expected from the BCS model [42]. In addition, the estimated in-plane CP is plotted in Fig. 5(b), which indicates that Se substitution in the examined samples results in an increase in in-plane CP [11]. See Ref [11] for details on the calculation of in-plane CP in LaO$_{0.5}$F$_{0.5}$BiS$_{2-x}$Se$_x$.

*3-2. Structural Properties*

The crystal structure parameters were investigated using Rietveld refinements, where the SXRD pattern and the refinement result at $x = 0.8$ ($T = 301$ K) are displayed in Fig. 3. See the Supplemental Materials [43] for the other compositions. LaF$_3$ impurities were included in all refinements; however, the population of the phase was minor (1–2%). The estimated compositions (Se concentration $x$) were close to nominal $x$ and those determined by EDX, as listed in Table I. The temperature dependence of the estimated lattice constants ($a$ and $c$) and the cell volume ($V$) are plotted in Figs. 4(a–c) at $x = 0.8$ and in Supplemental Materials [43] for other compositions. The temperature differential of $V$ was estimated by linear fitting of the $V$–$T$ data, and $\beta_V$ was calculated using equation (2). The $v_L$ was estimated as 3110(10), 3260(10), 3290(10), 3290(10), and 3290(10) m/s at $x = 0.2, 0.4, 0.6, 0.8$, and 1.0, respectively. An example of fitting the specific heat data above $T_c$ is shown in Fig. 4(d). The estimated $\theta_D$ is 220(1), 215(1), 221(1), 218(1), 221(1) K at $x = 0.2, 0.4, 0.6, 0.8$, and 1.0, respectively. Here, we analyzed low-temperature specific heat data at 3–9 K for $x = 0$ by refining fitting and obtained 228(1) K (See Fig. S6 in Supplemental Materials [43] for the analysis for $x = 0$). Therefore, the $B$ and $\gamma_G$ are modified from the published data in Ref. [28]. Using equations (1–5), we calculated $B$ and $\gamma_G$ and plotted them in Figs. 5(c and d). $B$ increased with increasing in-plane CP in REO$_{0.5}$F$_{0.5}$BiS$_2$ (RE = La, Ce, Pr, Nd). In LaO$_{0.5}$F$_{0.5}$BiS$_{2-x}$Se$_x$ [28], $B$ increases with a small amount of Se and became the highest at $x = 0.4$. Subsequently, $B$ decreased with increasing $x$ at $x \geq 0.4$. Because the in-plane CP, shown in Fig. 5(b), increases with increasing $x$, the structural characteristics of REO$_{0.5}$F$_{0.5}$BiS$_2$ and LaO$_{0.5}$F$_{0.5}$BiS$_{2-x}$Se$_x$ differ. The $\gamma_G$ values for $x = 0$ and 0.2 are smaller than those for $x \geq 0.4$. However, it becomes relatively constant at $x \geq 0.4$. Therefore, bulk superconductivity in LaO$_{0.5}$F$_{0.5}$BiS$_{2-x}$Se$_x$ emerges when the anharmonicity is enhanced as seen with $x \geq 0.4$, which implies positive correlation between $\gamma_G$ and superconductivity. For $x = 0$ and 0.2, the trend looks inconsistent with the scenario, but we consider that the data point at $x = 0$ was slightly increased due to the presence of local structural disorder, where local-scale symmetry lowering is present [44]. See Fig. S7 in Supplemental Materials [43] for the estimated isotropic displacement parameter $U_{iso}$ for the Ch1 site, which clearly shows that $U_{iso}$ at $x = 0$ is large due to the local disorder. When focusing on $T_c$, however, the results did not reveal a clear positive correlation between $T_c$ and $\gamma_G$ in LaO$_{0.5}$F$_{0.5}$BiS$_{2-x}$Se$_x$. In the following discussion, we discuss the relationship between superconductivity and lattice anharmonicity in REO$_{0.5}$F$_{0.5}$BiCh$_2$.

## 4. Discussion

*4-1. Correlation between superconductivity and anharmonicity*

According to BCS theory, a high phonon energy is preferred for a higher $T_c$, as revealed in $MgB_2$ [45] and hydride superconductors [46–48]. The target system of this study is $BiCh_2$-based compounds, in which the Bi p electrons contribute to electronic conduction and, hence, superconducting pairing. Because of the heavy Bi atoms, a high phonon energy cannot be expected in $BiCh_2$-based compounds, and a theoretical study suggested that several-K $T_c$ in tetragonal $RE(O,F)BiS_2$ cannot be understood using conventional phonon-mediated models [25]. In addition, the absence of isotope effects on the S or Se sites has been confirmed in tetragonal $Bi_4O_4S_3$ and $La(O,F)BiSSe$ [16,18]. Therefore, the pairing mechanisms of superconductivity in tetragonal $BiCh_2$-based compounds are different from conventional phonon-mediated weak-coupling superconductivity. Here, we discuss the relationship between the anharmonic atomic vibrations and superconductivity in $BiCh_2$-based compounds. In caged compounds, the increase in anharmonicity in the rattling mode (low-energy optical-phonon mode) results in the enhancement of strong-coupling superconductivity and an increase in $T_c$ [33,49]. If the pairing in $BiCh_2$-based systems is mediated by the rattling of Bi, the absence of the isotope effect for the S or Se sites can be explained.

In Fig. 6, $T_c$s of $LaO_{0.5}F_{0.5}BiS_{2-x}Se_x$ and $REO_{0.5}F_{0.5}BiS_2$ (RE = Pr, Nd) are plotted as functions of $\gamma_G$. As mentioned above, no clear correlation between $T_c$ and $\gamma_G$ was observed for $LaO_{0.5}F_{0.5}BiS_{2-x}Se_x$, which could be due to the constant $\gamma_G$ at $x = 0.4$–1.0. However, we find that $T_c$ increases with increasing $\gamma_G$ for Se-free and Se-poor samples, as highlighted by the green curve in Fig. 6, after the addition of the data points of $REO_{0.5}F_{0.5}BiS_2$. Here, we excluded $x = 0$ because the susceptibility does not show bulk superconductivity signals. What is the origin of the large $\gamma_G$, which suggests large anharmonicity in the system? We consider that the one-dimensional rattling-like vibrations of Bi atoms along the *c*-axis contribute to their anharmonicity. Indeed, the softening of these rattling modes was detected using $\gamma_G$ in thermoelectric $LaOBi(S,Se)_2$ [29,32]. Therefore, the increase in $\gamma_G$ in superconducting $REO_{0.5}F_{0.5}BiCh_2$ with Se-poor compositions is possibly linked to softening of the low-energy phonon mode of Bi in the $BiCh_2$ layers. In the current work, we cannot clarify the cause of the deviation in the data points for the Se-rich samples in Fig. 6, but it would be related to the changes in superconducting states (and/or the influence of anharmonicity on superconductivity) because Se substitution at the Ch1 site largely modifies the chemical bonding and electronic states [10,15]. In addition, we discuss possible changes in characteristics of anharmonic vibration by Se substitution.

*4-2. Low-energy phonons*

Because the discussion above assumes an anharmonic rattling-like vibration in superconducting REO$_{0.5}$F$_{0.5}$BiCh$_2$, we examined the presence of low-energy optical phonons in these samples using the specific heat. We performed low-temperature specific heat measurements for LaO$_{0.5}$F$_{0.5}$BiS$_{2-x}$Se$_x$ and LaOBiS$_{2-y}$Se$_y$ (thermoelectric phase). In materials with low-energy optical phonons, a peak is expected in the plot at a temperature corresponding to approximately a fifth of the phonon energy [50,51]. Because the phonon energy observed in inelastic neutron scattering (INS) is 6–7 meV for LaOBiS$_{2-y}$Se$_y$ [32], we expected a peak at approximately 12–14 K. As shown in Fig. S8(a) (Supplemental Materials [43]), the specific heat analysis results for LaOBiS$_{2-y}$Se$_y$ confirmed the evolution of a peak structure and the softening of the optical phonon [decrease in peak temperature ($T_p$)] by Se substitution, which is consistent with the INS results [32]. The analysis results for LaO$_{0.5}$F$_{0.5}$BiS$_{2-x}$Se$_x$ are summarized in Fig. 7(a), where $(C-\gamma T)/T^3$ is plotted as functions of temperature. Similar peaks were observed at 12.4–13.1 K for LaO$_{0.5}$F$_{0.5}$BiS$_{2-x}$Se$_x$. The peak temperatures were determined by taking the temperature differential of the polynomial-fitting results. The peaks are evidence of low-energy phonon modes in superconducting LaO$_{0.5}$F$_{0.5}$BiS$_{2-x}$Se$_x$. In addition, the peak temperatures for LaO$_{0.5}$F$_{0.5}$BiS$_{2-x}$Se$_x$ are clearly lower than $T_p$ = 14.0 K for LaOBiS$_2$ ($y$ = 0) [see Fig. S8(b)], implying the enhanced anharmonicity by F substitution. Based on the difference in $\gamma_G$, we expected a lower $T_p$ at $x$ = 0.8 than that at $x$ = 0.2, but the obtained results did not corroborate this expectation. Although we have no explanation for this trend, one possible cause is the difference in local structural disorder in samples with lower in-plane CP (e.g., at $x$ = 0 and 0.2). The change in local structure may affect $T_p$. Furthermore, from Fig. 7(b), we find that the optical phonon energy decreases with Se substitution for $x \leq 0.4$, but it increases for further Se substitution for $x$ = 0.6–1.0. However, we noticed that the contributions of optical phonon in $(C-\gamma T)/T^3$ is large at $x$ = 0.8 and 1.0. The large $(C-\gamma T)/T^3$ values than $\beta$ (see Table I) is linked to the enhanced anharmonicity as revealed in Ref. 51. Therefore, the weak hardening at $x$ = 0.6–1.0 would be related to the changes in the shape of the interatomic potential and/or optical phonon characteristics. The results on $\gamma_G$ and $(C-\gamma T)/T^3$ would be suggesting the importance of anharmonicity on superconductivity in LaO$_{0.5}$F$_{0.5}$BiS$_{2-x}$Se$_x$. To further understand the correlation between superconductivity and anharmonicity (rattling-like Bi vibration) in BiCh$_2$-based superconductors, further studies using various probes, including INS, are required.

## 5. Conclusion

Polycrystalline samples of LaO$_{0.5}$F$_{0.5}$BiS$_{2-x}$Se$_x$ were synthesized. Using magnetic susceptibility and specific heat measurements, the superconducting properties of LaO$_{0.5}$F$_{0.5}$BiS$_{2-x}$Se$_x$ were examined, and the $T_c$ phase diagram obtained. To investigate the lattice anharmonicity, the Grüneisen parameter ($\gamma_G$) was calculated using the sample density; sound velocity; Debye temperature; and crystal structure parameters, including the volume thermal expansion coefficient and specific heat. A larger $\gamma_G$ value was observed at $x \geq 0.4$, at which $T_c$ was higher than that at $x = 0.2$, and bulk superconductivity was observed. In addition, the discussion including REO$_{0.5}$F$_{0.5}$BiS$_2$ (RE = Pr, Nd) showed that $T_c$ increased with increasing $\gamma_G$ for samples with Se-free and Se-poor compositions. The trend suggests positive correlation between anharmonicity and superconductivity in the Se-poor samples. However, the data for the Se-rich samples deviate from the $\gamma_G$-$T_c$ correlation. These results imply that superconducting states and/or the influence of anharmonicity on superconductivity are different between the Se-poor and Se-rich phases of REO$_{0.5}$F$_{0.5}$BiCh$_2$. To reveal the origin of the anharmonicity, we measured the specific heat and observed a peak in $(C-\gamma T)/T^3$ at 12.4–13.1 K for LaO$_{0.5}$F$_{0.5}$BiS$_{2-x}$Se$_x$. Those results suggest the importance of anharmonic vibration on superconductivity in LaO$_{0.5}$F$_{0.5}$BiS$_{2-x}$Se$_x$.

**Acknowledgment**

The authors thank O. Miura for supports in susceptibility experiments. This work was partially supported by Grant-in-Aid for Scientific Research (KAKENHI) (Nos. 18KK0076, 21H00151), JST-CREST (No. JPMJCR20Q4), and Tokyo Metropolitan Government Advanced Research (No. H31-1).

*E-mail:mizugu@tmu.ac.jp

Table I. Parameters estimated from the composition and crystal structure analyses and the physical property measurements.

| Nominal $x$ | 0.2 | 0.4 | 0.6 | 0.8 | 1.0 |
|---|---|---|---|---|---|
| $x$ (EDX) | 0.131(8) | 0.40(2) | 0.600(11) | 0.778(10) | 0.969(14) |
| Refined $x$ | 0.16(2) | 0.44(2) | 0.61(2) | 0.87(2) | 1.00(2) |
| Space group | Tetragonal $P4/nmm$ (#129) | | | | |
| Relative density (%) | 96 | 96 | 96 | 98 | 98 |
| $a$ (Å) ($T \sim 300$ K) | 4.07513(5) | 4.08667(4) | 4.10009(6) | 4.11241(4) | 4.12604(5) |
| $c$ (Å) ($T \sim 300$ K) | 13.3792(2) | 13.4101(2) | 13.4367(3) | 13.4622(2) | 13.5116(2) |
| $V$ (Å$^3$) ($T \sim 300$ K) | 222.184(5) | 223.960(5) | 225.882(6) | 227.672(4) | 230.024(6) |
| In-plane CP | 1.015 | 1.021 | 1.026 | 1.031 | 1.035 |
| $\beta_V$ (µK$^{-1}$) | 35.8(4) | 35.9(6) | 39.6(6) | 35.8(6) | 36.7(5) |
| $T_c$ (K) ($\chi$) | 3.10 | 3.25 | 3.50 | 3.80 | 3.65 |
| $T_c$ (K) ($C$) | 2.99 | 3.25 | 3.27 | 3.58 | 3.48 |
| $\Delta C/\gamma T_c$ | 0.53 | 0.56 | 0.57 | 0.72 | 1.18 |
| $\gamma$ (mJ/mol K$^2$) | 3.8 | 4.1 | 4.4 | 4.5 | 3.4 |
| $\beta$ (mJ/mol K$^4$) | 0.915(10) | 0.975(13) | 0.904(13) | 0.936(11) | 0.903(13) |
| $\theta_D$ (K) | 220(1) | 215(1) | 221(1) | 218(1) | 221(1) |
| $V_L$ (m/s) | 3110(10) | 3260(10) | 3290(10) | 3290(10) | 3290(10) |
| $C_V$ | 124.61 mJ/mol K (theoretical value) | | | | |
| $B$ (GPa) | 38.9(3) | 47.2(3) | 46.6(3) | 45.0(3) | 44.3(3) |
| $\gamma_G$ | 0.752(14) | 0.92(2) | 0.93(2) | 0.89(2) | 0.92(2) |

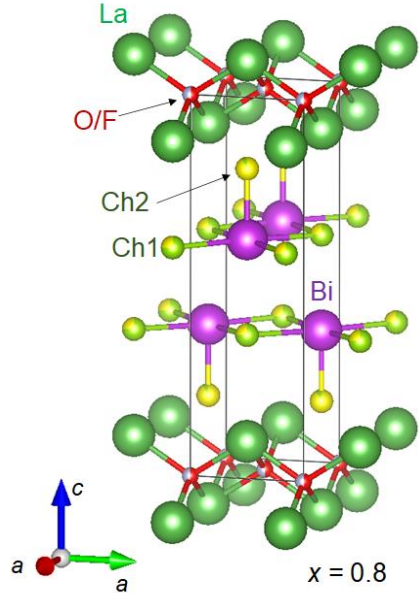

Fig. 1. (Color Online) Schematic image of crystal structure of LaO$_{0.5}$F$_{0.5}$BiS$_{1.2}$Se$_{0.8}$ ($x = 0.8$). The solid square indicates a unit cell.

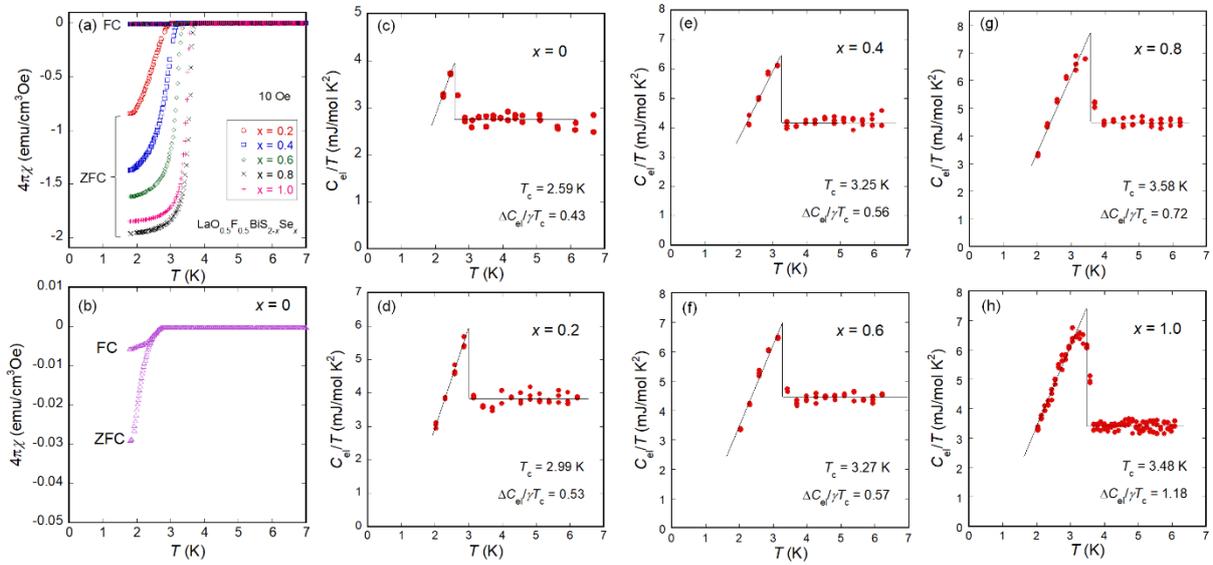

Fig. 2. (Color Online) Superconducting properties of the LaO$_{0.5}$F$_{0.5}$BiS$_{2-x}$Se$_x$ samples with $x = 0.2, 0.4, 0.6, 0.8$, and 1.0. (a,b) Temperature dependences of ZFC and FC magnetic susceptibility ($4\pi\chi$). (c–h) Temperature dependences of electronic specific heat ($C_{el}$) in a form of $C_{el}/T$. In the plots, the $C_{el}$ jump ($\Delta C_{el}$) was estimated by solid lines after considering entropy balance at the superconducting transition. The asymmetric evolution of the specific heat at around $T_c$ would be due to the affection of local structural disorder that essentially exist in this system (Fig. S7).

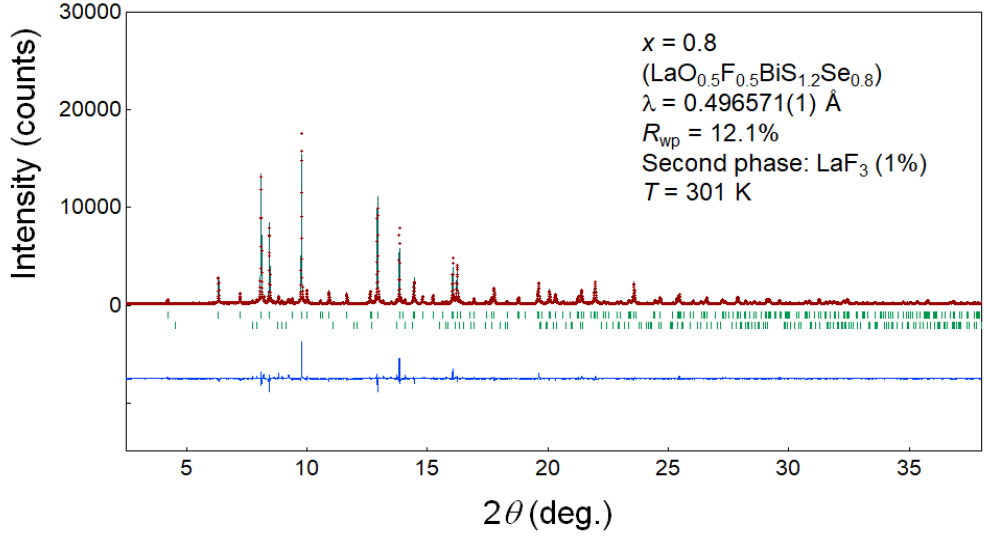

Fig. 3. (Color Online) SXRD pattern and Rietveld refinement result for $x = 0.8$.

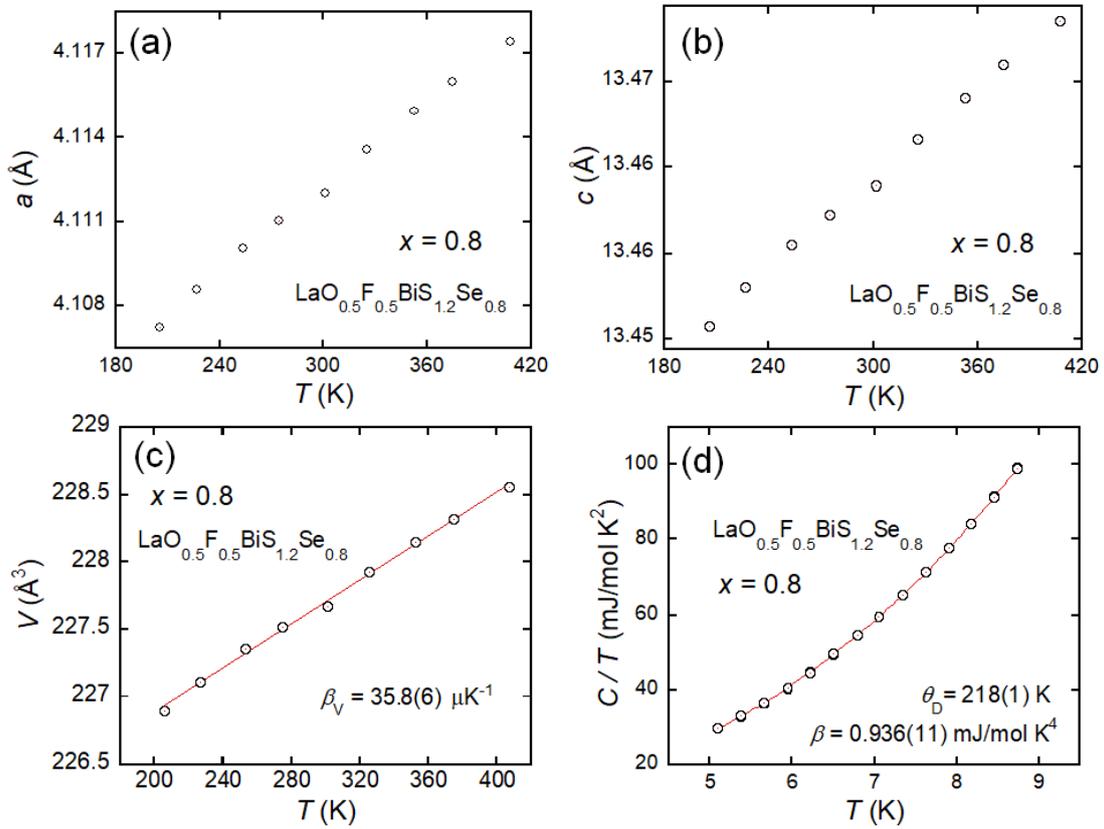

Fig. 4. (Color Online) (a–c) Temperature dependences of the lattice constant ($a$, $c$) and the cell volume ($V$) for $x = 0.8$. The volume thermal expansion coefficient ($\beta_V$) was estimated from linear fitting of $V$-$T$ data, and the estimated $\beta_V$ is displayed. (d) Fitting result of $C$ using $C = \gamma T + \beta T^3 + \delta T^5$. The estimated $\beta$ and the Debye temperature ($\theta_D$) are displayed. See Supplemental Materials [43] for analysis results for other $x$.

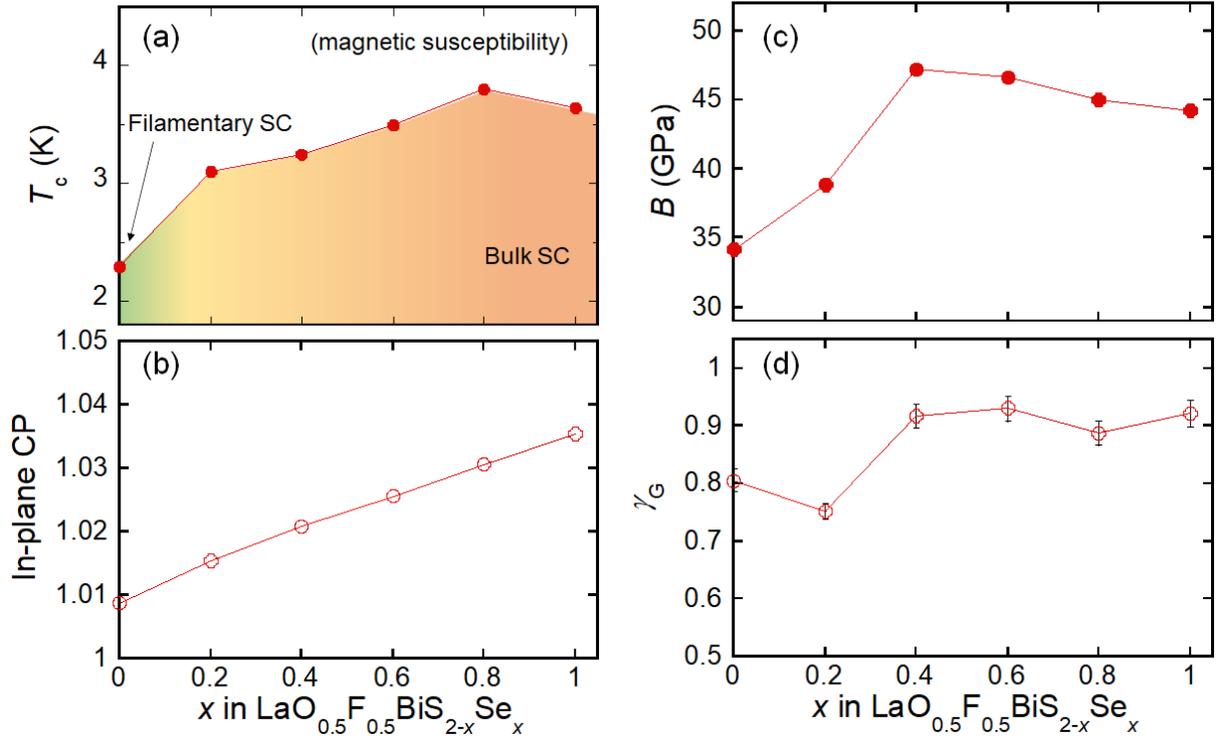

Fig. 5. (Color Online) Se concentration ($x$) dependences of (a) $T_c$ (magnetic susceptibility), (b) in-plane chemical pressure (CP) estimated by the same calculation in Ref. 11, (c) bulk modulus ($B$), and (d) Grüneisen parameter ($\gamma_G$). The data for $x = 0$ were modified from our published parameter by modifying Debye temperature ($\theta_D = 228$ K).

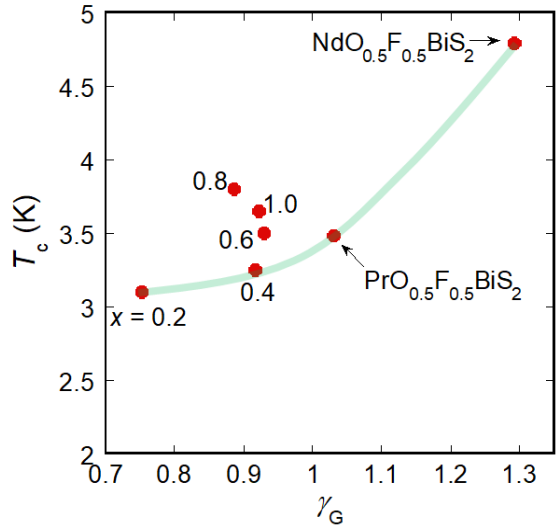

Fig. 6. (Color Online) $T_c$ plotted as a function of Grüneisen parameter ($\gamma_G$) of REO$_{0.5}$F$_{0.5}$BiCh$_2$. The green curve is eye-guide. $x$ indicates Se concentration in LaO$_{0.5}$F$_{0.5}$BiS$_{2-x}$Se$_x$.

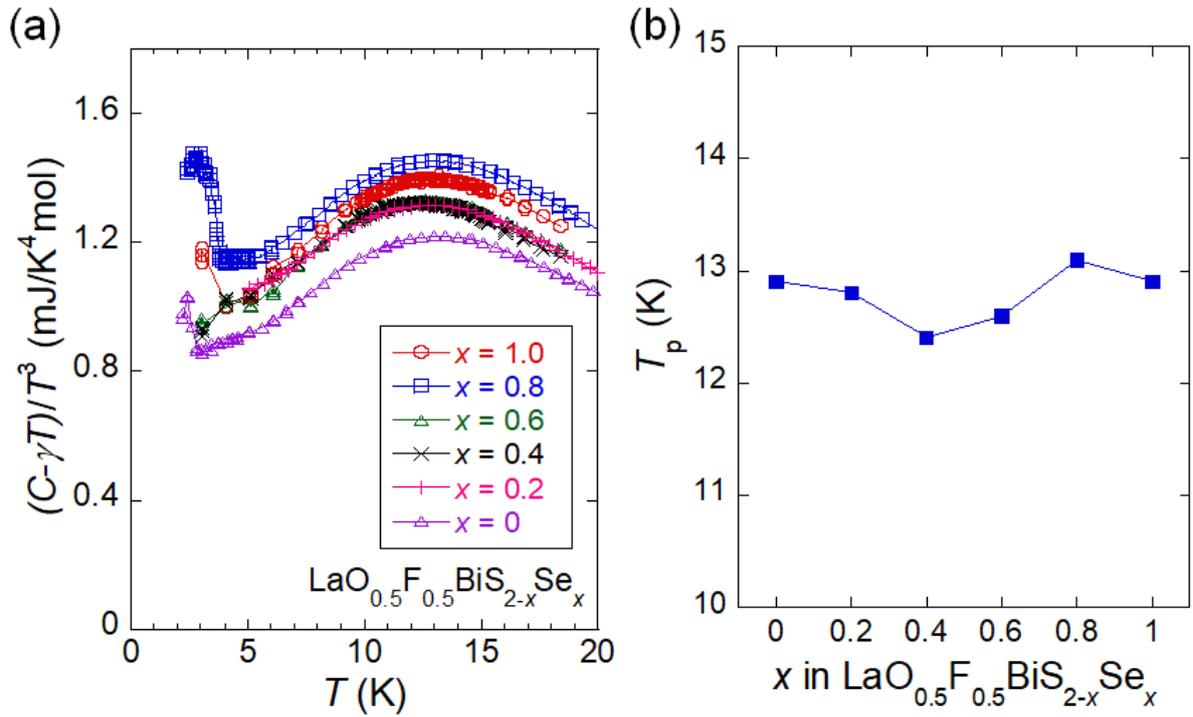

Fig. 7. (Color Online) (a) Temperature dependences of optical phonon contribution in $C$ [$(C-\gamma T)/T^3$ or $C/T^3$] for $LaO_{0.5}F_{0.5}BiS_{2-x}Se_x$. (b) Se concentration dependence of peak temperature ($T_p$), which was determined by polynomial fitting of the data shown in (a).

Supplemental Materials

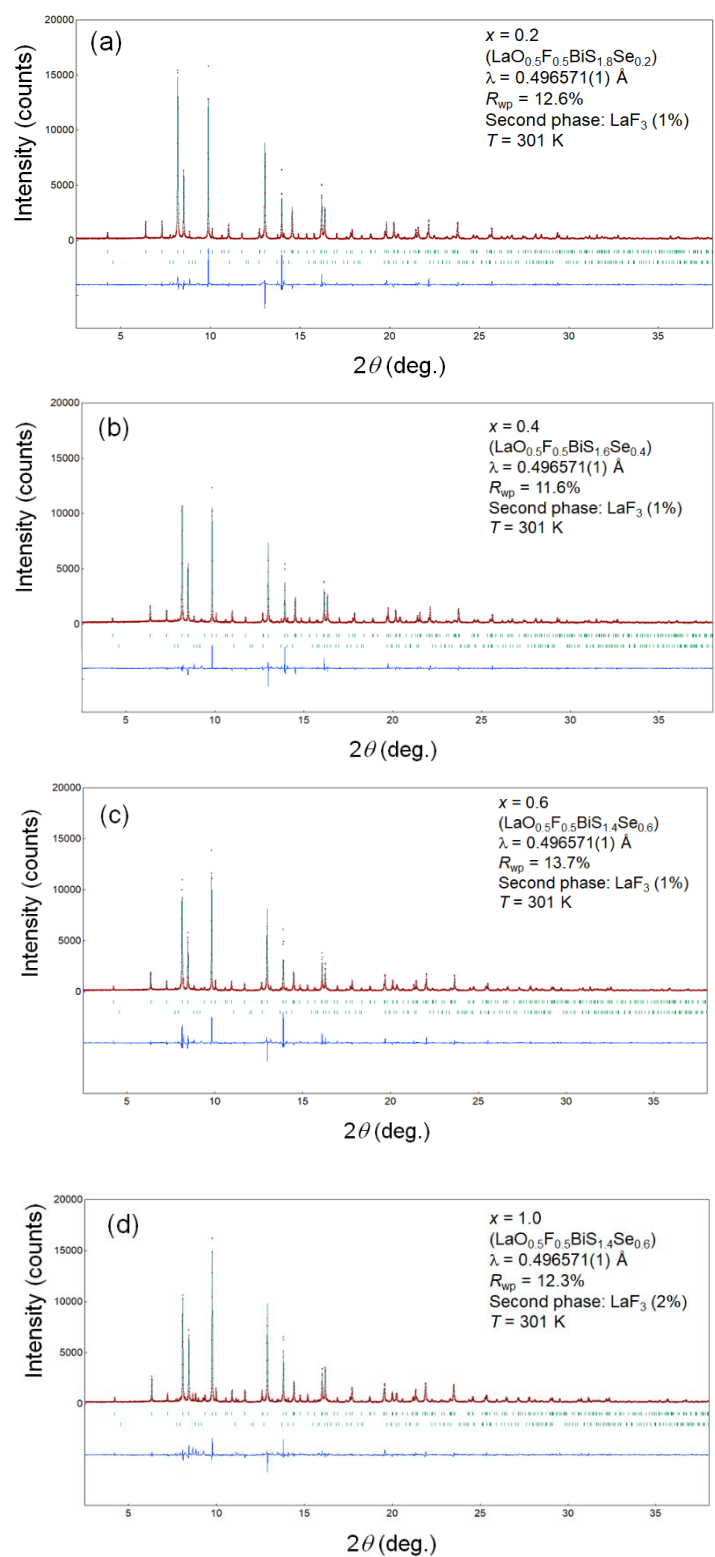

Fig. S1. Synchrotron X-ray diffraction patterns and Rietveld refinement results for $LaO_{0.5}F_{0.5}BiS_{2-x}Se_x$ [(a) $x = 0.2$, (b) $x = 0.4$, (c) $x = 0.6$, (d) $x = 1.0$]. See main text for the result for $x = 0.8$.

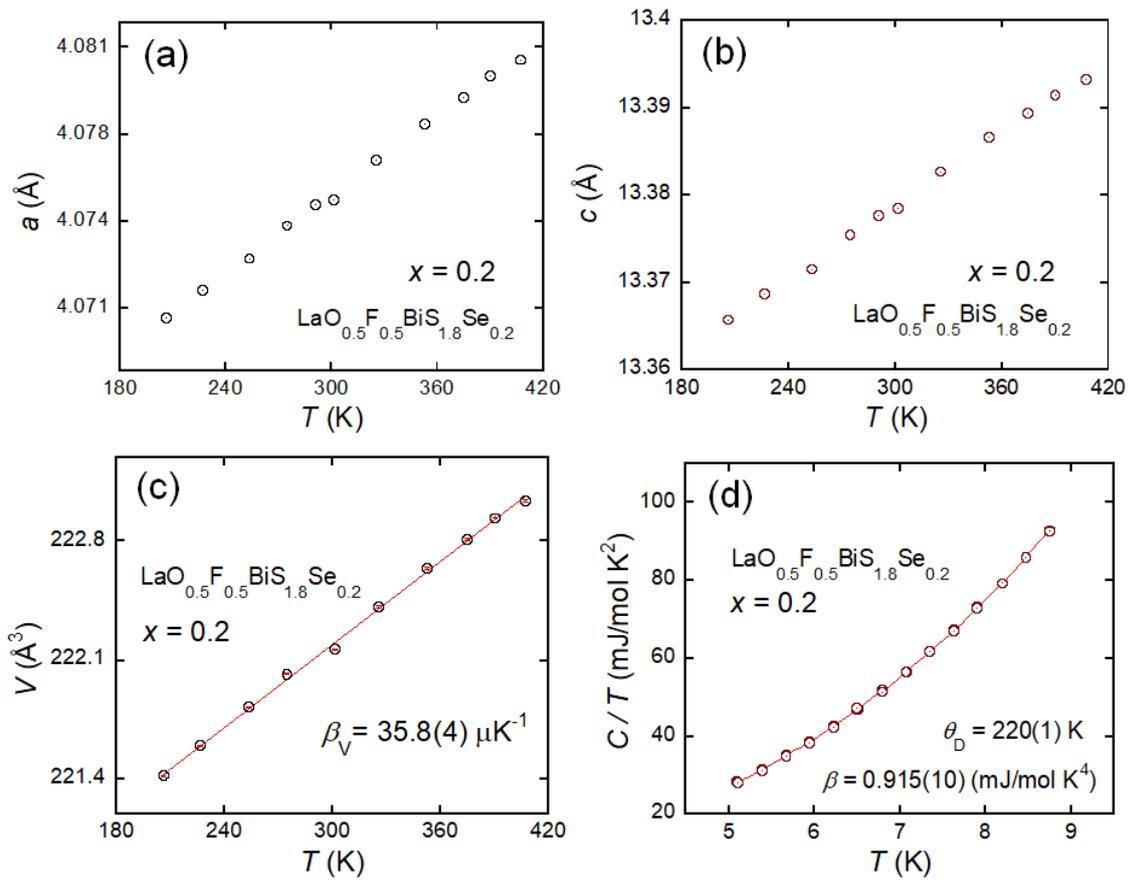

Fig. S2. (a–c) Temperature dependences of the lattice constant ($a$, $c$) and the cell volume ($V$) for $x = 0.2$. The volume thermal expansion coefficient ($\beta_V$) was estimated from linear fitting of $V$-$T$ data, and the estimated $\beta_V$ is displayed. (d) Fitting result of $C$ using $C = \gamma T + \beta T^3 + \delta T^5$. The estimated $\beta$ and the Debye temperature ($\theta_D$) are displayed.

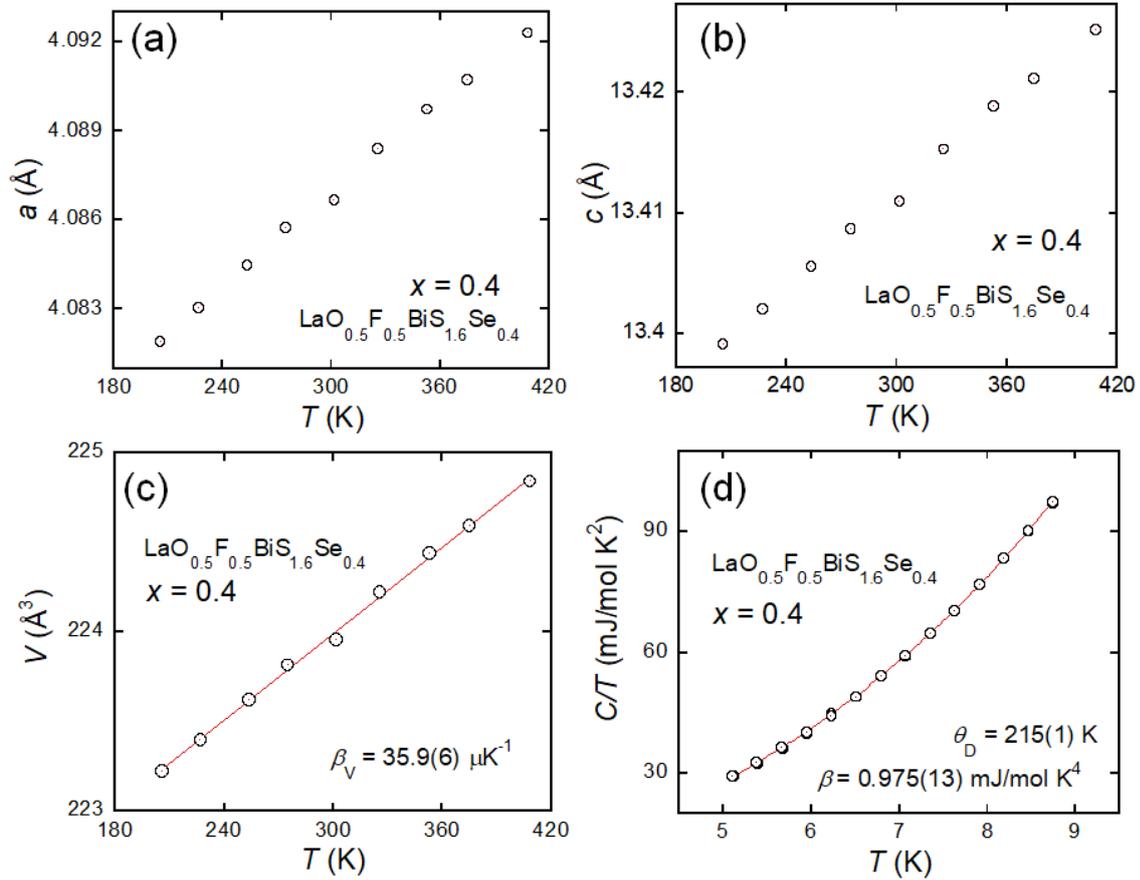

Fig. S3. (a–c) Temperature dependences of the lattice constant ($a$, $c$) and the cell volume ($V$) for $x = 0.4$. The volume thermal expansion coefficient ($\beta_V$) was estimated from linear fitting of $V$-$T$ data, and the estimated $\beta_V$ is displayed. (d) Fitting result of $C$ using $C = \gamma T + \beta T^3 + \delta T^5$. The estimated $\beta$ and the Debye temperature ($\theta_D$) are displayed.

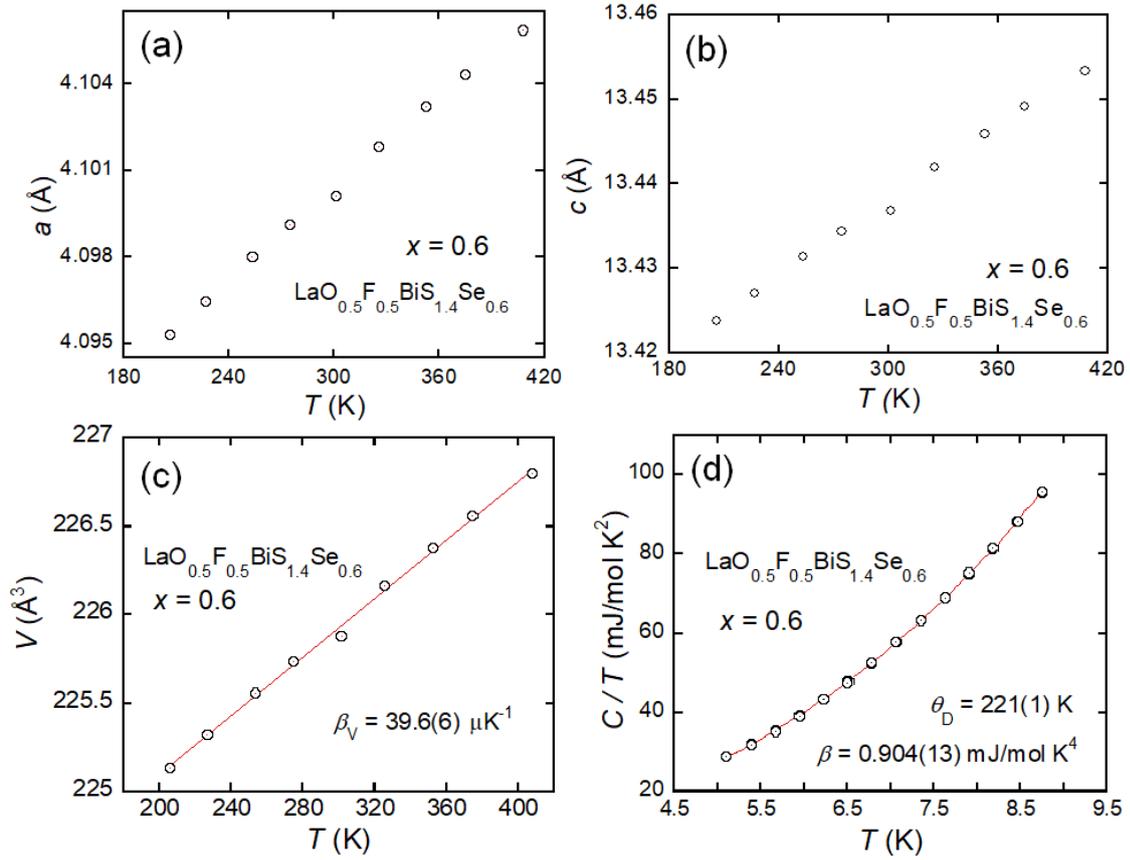

Fig. S4. (a–c) Temperature dependences of the lattice constant ($a$, $c$) and the cell volume ($V$) for $x$ = 0.6. The volume thermal expansion coefficient ($\beta_V$) was estimated from linear fitting of $V$-$T$ data, and the estimated $\beta_V$ is displayed. (d) Fitting result of $C$ using $C = \gamma T + \beta T^3 + \delta T^5$. The estimated $\beta$ and the Debye temperature ($\theta_D$) are displayed.

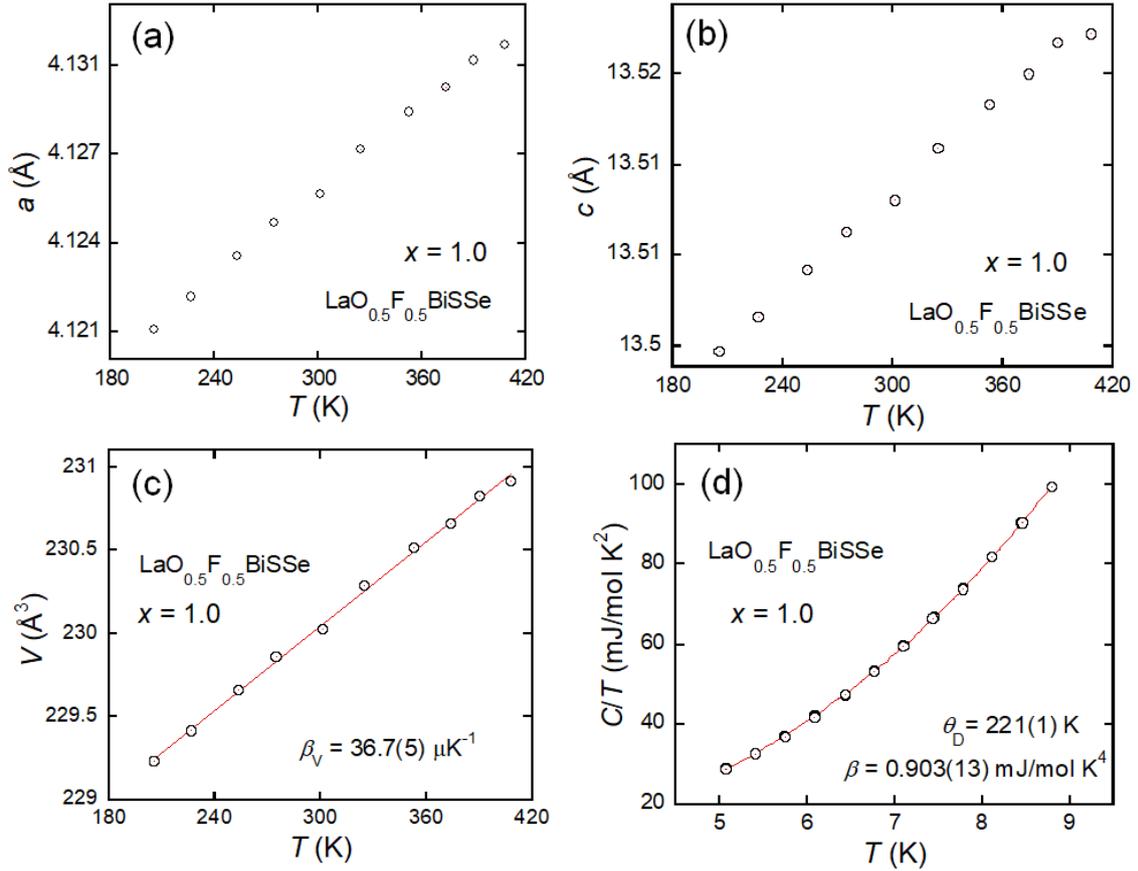

Fig. S5. (a–c) Temperature dependences of the lattice constant ($a$, $c$) and the cell volume ($V$) for $x = 1.0$. The volume thermal expansion coefficient ($\beta_V$) was estimated from linear fitting of $V$-$T$ data, and the estimated $\beta_V$ is displayed. (d) Fitting result of $C$ using $C = \gamma T + \beta T^3 + \delta T^5$. The estimated $\beta$ and the Debye temperature ($\theta_D$) are displayed.

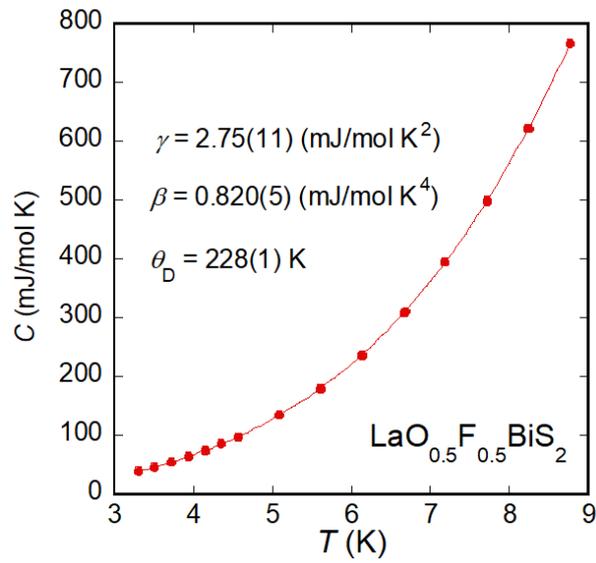

Fig. S6. Low-temperature specific heat analysis result for $x = 0$.

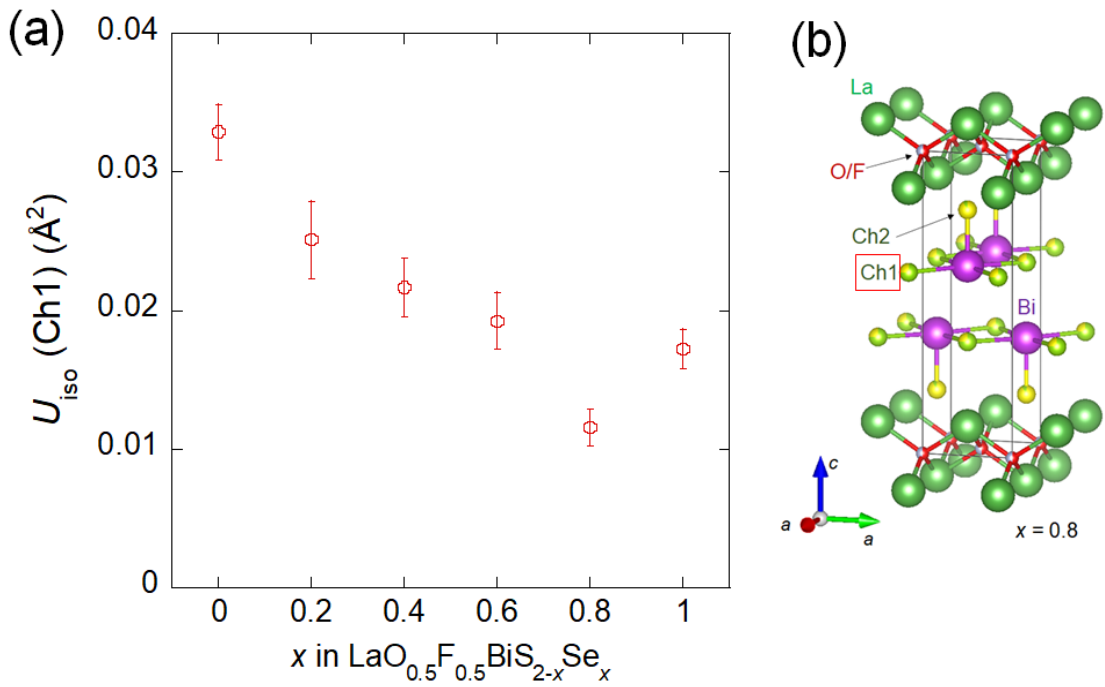

Fig. S7. (a) Nominal $x$ dependence of isotropic displacement parameter $U_{iso}$ for the Ch1 site. (b) Schematic image of the crystal structure and sites of $LaO_{0.5}F_{0.5}BiS_{2-x}Se_x$ (with structural parameters of $x = 0.8$).

[LaOBiS$_{2-y}$Se$_y$]

In this work, the obtained results on LaO$_{0.5}$F$_{0.5}$BiS$_{2-x}$Se$_x$ were discussed with data for thermoelectric phases LaOBiS$_{2-y}$Se$_y$. The polycrystalline samples were prepared by solid state reaction as described in Ref. S1.

[S1] F. I. Abbas, A. Yamashita, K. Hoshi, R. Kiyama, Md. R. Kasem, Y. Goto, and Y. Mizuguchi, Appl. Phys. Express **14**, 071002 (2021).

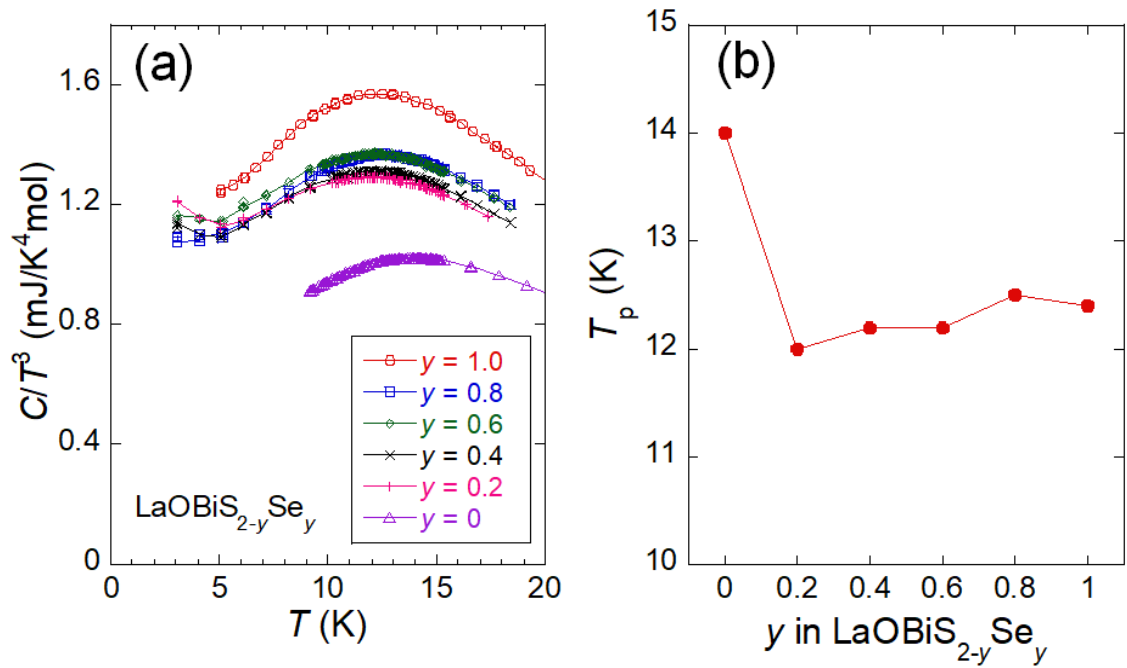

Fig. S8. (a) Temperature dependence of $C/T^3$ for LaOBiS$_{2-y}$Se$_y$. (b) Se concentration dependence of peak temperature ($T_p$).

Table S1. Obtained $\beta$ and $\theta_D$ for LaOBiS$_{2-y}$Se$_y$. $\beta$ was determined by analyzing low-temperature specific heat with a formula of $C = \beta T^3 + \delta T^5$.

| $y$ | 0 [S1] | 0.2 | 0.4 | 0.6 | 0.8 | 1.0 [S1] |
|---|---|---|---|---|---|---|
| $\beta$ (mJ/K$^4$mol) | 0.883(4) | 1.118(6) | 1.041(6) | 1.099(4) | 1.024(4) | 0.946(3) |
| $\theta_D$ (K) | 223 | 205 | 211 | 207 | 212 | 218 |